\let\accentvec\vec
\documentclass[runningheads]{llncs}

\let\vec\accentvec
\usepackage{amsmath}
\usepackage{graphicx}
\usepackage{multirow}
\usepackage[caption=false]{subfig}
\usepackage{ulem}
\usepackage{cancel}
\usepackage{dsfont} 
\usepackage{bbm,epsfig,graphics,epic,color,rotating,color}

\DeclareMathOperator*{\scale}{{\bf scale}}
\DeclareMathOperator*{\diag}{diag}

\def\ci{\perp\!\!\!\perp}
\makeatletter
\renewcommand\section{\@startsection{section}{1}{\z@}%
	{-8\p@ \@plus -4\p@ \@minus -4\p@}%
	{6\p@ \@plus 4\p@ \@minus 4\p@}%
	{\normalfont\large\bfseries\boldmath
		\rightskip=\z@ \@plus 8em\pretolerance=10000 }}
\renewcommand\subsection{\@startsection{subsection}{2}{\z@}%
	{-8\p@ \@plus -4\p@ \@minus -4\p@}%
	{6\p@ \@plus 4\p@ \@minus 4\p@}%
	{\normalfont\normalsize\bfseries\boldmath
		\rightskip=\z@ \@plus 8em\pretolerance=10000 }}
\renewcommand\subsubsection{\@startsection{subsubsection}{3}{\z@}%
	{-3\p@ \@plus -4\p@ \@minus -4\p@}%
	{-1.5em \@plus -0.22em \@minus -0.1em}%
	{\normalfont\normalsize\bfseries\boldmath}}
\makeatother

\begin{document}
%

\title{Network reconstruction with local partial
	correlation: a comparative evaluation}
%
\titlerunning{Network reconstruction with local partial
	correlation}
%
\author{Henrique Bolfarine\thanks{The corresponding author received support by CNPq grant 133935/2015-9.}  \and Lina Thomas\thanks{The author received support by FAPESP grant 2013/06223-1.} \and Anatoly Yambartsev\thanks{The author thanks CNPq grant 301050/2016-3.}}
\authorrunning{H. Bolfarine et al.}
%
\institute{Instituto de Matem\'atica e Estat\'istica, \\ Universidade de S\~ao Paulo, S\~ao Paulo-SP, Brazil\\ \email{bolfarin@ime.usp.br}}
\maketitle              
\begin{abstract}
Over the past decade, various methods have been proposed for the reconstruction of networks modeled as Gaussian Graphical Models. In this work, we analyzed three different approaches: the Graphical Lasso (GLasso), the 
Graphical Ridge (GGMridge), and the Local Partial Correlation (LPC). For the evaluation of the methods, we used high dimensional data generated from simulated random graphs (Erd\"os-R\'enyi, Barab\'asi-Albert, Watts-Strogatz). The performance was assessed through the Receiver Operating Characteristic (ROC) curve. In addition, the methods were used to reconstruct the co-expression network for differentially expressed genes in human cervical cancer data. The
LPC method outperformed the GLasso in most simulated cases. The GGMridge produced better ROC curves then both the other methods. Finally, LPC and GGMridge obtained similar outcomes in real data studies.

\keywords{network reconstruction, Gaussian Graphical Model, partial correlation, gene co-expression, regularization.}
\end{abstract}
\section{Introduction}

The reconstruction of network structures through estimated associations has become more popular over the past decade, mainly due to the availability of massive data sets. 
Several methods of network reconstruction have been recently developed. Most of these methods are based on Graphical Models (GMs) \cite{GM_book} due to their ability to represent conditional dependencies over graph structures. If the variables in these models are assumed to have Gaussian distribution, then we have a Gaussian Graphical Model (GGM), which facilitates the use of partial correlation to identify conditional dependence \cite{Mardia}. The main problem that emerges during network reconstruction on large data sets is that in some cases the number of variables is considerably larger than the sample size.

To overcome this problem, \cite{LFY2012} and \cite{TVSYM2016} proposed the Local Partial Correlation method (LPC). The method estimates a partial correlation between two variables using the neighborhood of the relevance network \cite{schafer2005}, composed of the highest Pearson correlation of both variables.  
Our objective relies on assessing  the  efficiency of LPC by comparing it with two other approaches based on regularization methodologies: the GLasso and the GGMridge. GLasso, defined in \cite{friedman2008sparse}, is the most popular method, and it is based on LASSO \cite{tibshirani1996regression}, a technique widely applied in regression analysis, mainly for variable and model selection. GGMridge \cite{ha2014partial}, based on the work of \cite{schafer2005empirical}, estimates the partial correlation matrix using Ridge penalty and performs statistical estimation using empirical distributions. 

The assessment of the methods was performed using simulations and real data. Regarding the simulation studies, we compared the performance of the methods by generating high dimensional data from the following random graph structures: Erd\"os-R\'enyi, a well-known uniform model \cite{erdos1960evolution}, Barab\'asi-Albert or scale-free model \cite{barabasi1999emergence}, frequently used to model biological interactions, and Watts-Strogatz or small-world \cite{watts1998collective}, a popular model for social interactions. The ROC curves were employed to compare the performance of each method, where there are original network structures to liken. 

The real data came from \cite{zhai2007gene}, which is composed of gene expressions from tumorous cervical cancer cells. From this original data set, we selected 1268 genes known as differentially expressed genes (DEGs), identified by \cite{mine2013gene}. We applied the methods in this data set and compared the results of the reconstructions in terms of the number of nodes and edges that were identified in common by the methods.

Before presenting the different approaches in Section \ref{secLPC}, and the results in Section \ref{sim} and Section \ref{gene_express}, we shall recall basic notions of GGM.

\section{Methods}\label{secLPC}

\subsection{Gaussian Graphical Models}\label{secGGM}


Let $X_{\cal V} = (X_1, \dots, X_p) \in \bbbr^p$ be a $p$-dimensional random vector, with ${\cal V} = \{1,\dots, p\}$ and covariance matrix $\mathbf{\Sigma}_\mathcal{V}$, and let ${\cal G} = ({\cal V}, {\cal E})$ be a finite graph with set of vertices ${\cal V}$ and set of edges ${\cal E}\subseteq {\cal V}\times {\cal V}$. If $X_{\cal V}$ has Gaussian distribution and $\mathbf{\Sigma}_\mathcal{V}$ is positive definite, then $\mathbf{\Omega} \equiv (\omega_{ij})_{i,j\in\cal{V}} = \mathbf{\Sigma}^{-1}_\mathcal{V}$ is the precision matrix. 

Under multivariate Gaussian distribution assumption, if $\omega_{ij}=0$ , then the partial correlation
\begin{equation}\label{eq1}
\rho_{ij.{\cal Z}}=-\frac{\omega_{ij}}{\sqrt{\omega_{ii}\omega_{jj}}}\quad\text{for}\quad  i,j\in {\cal V},\quad\text{and} \quad{\cal Z}={\cal V}\setminus \{i,j\},
\end{equation}
between $X_i$ and $X_j$ is also zero, which implies conditional independence of these two variables given the rest ($X_i\ci X_j|X_{{\cal Z}}$) \cite{GM_book}. In the GGM context, this is equivalent to $\{i,j\} \notin {\cal E}$. Therefore, we can reconstruct a gene co-expression network by identifying if the elements of the precision and partial correlation matrices are different from zero. In summary, we have that
\begin{equation}\label{eq3}
\omega_{ij}=0 \Leftrightarrow \rho_{ij.{\cal Z}}=0 \Leftrightarrow X_i\ci X_j| X_{{\cal Z}}\Leftrightarrow \{i,j\}\notin{\cal E},\quad\text{for}\quad i,j \in {\cal V},
\end{equation}

\subsection{Local Partial Correlation}

Let $\mathbf{X}$ be a data matrix with $p$ variables and sample size $n$, and assume $p\gg n$ (high dimensional problem). For any fixed $i \in \mathcal{V}$ and p-value threshold $\alpha\in(0,1)$, define $\mathcal{Z}_{\alpha}(i) \subset\mathcal{V}\setminus \{i\}$ the set of indices of the variables that are significantly non-zero Pearson correlated with the variable $X_i$.
For any $i,j\in \mathcal{V}$, we define the neighborhood of $\{i, j\}$ the set $\mathcal{Z}_{\alpha}(i) \cup \mathcal{Z}_{\alpha}(j)$. If the neighborhood has more variables than samples, we select $\lfloor n/2 \rfloor$ variables from $\mathcal{Z}_{\alpha}(i) \cup \mathcal{Z}_{\alpha}(j)$ with the highest Pearson correlations.  

Denote $\mathcal{Z}_\alpha(i,j)\equiv \mathcal{Z}_{\alpha}(i) \cup \mathcal{Z}_{\alpha}(j)$ and $\mathcal{J} \equiv \{i,j\}\cup\mathcal{Z}_\alpha(i,j)$. We construct an empirical covariance matrix $\hat{\bf{\Sigma}}_{\cal J}$ that if inverted ($\hat{\bf{\Omega}}_{\cal J} ^{-1} $) provides the estimate $\hat{\rho}_{ij.{\cal Z}^{*}}$, where $\mathcal{Z}^{*}\equiv \mathcal{Z}_\alpha(i,j)$. Finally, we test $H_0:\rho_{ij.{\cal Z}^{*}}=0$  versus $H_a:\rho_{ij.{\cal Z}^{*}}\neq 0$ using the transformation
\begin{equation*}\label{eq:test_01}
\psi(\hat{\rho}_{ij.{\cal Z}^{*}})=\frac{1}{2}\log\{(1+\hat{\rho}_{ij.{\cal Z}^{*}})/(1-\hat{\rho}_{ij.{\cal Z}^{*}})\}.
\end{equation*}
For a significance level $\alpha_{LPC}\in (0,1)$, the null hypothesis is rejected if
\begin{equation}\label{eq2}
\sqrt{n-|{\cal Z}^{*}|-3}\times|\psi(\hat{\rho}_{ij.{\cal Z}^{*}})|>\Phi^{-1}(1-\alpha_{LPC}/2),
\end{equation}
where $\Phi(x)$ is the cumulative Gaussian distribution ${\cal N}(0,1)$, and $|{\cal Z}^{*}|$ is the size of the neighborhood. If (\ref{eq2}) is true, then from (\ref{eq3}), $\{i,j\}\in {\cal E}$.

\subsection{GLasso and GMMridge}

The constrained or penalized Maximum Likelihood Estimate (MLE) is often used for high dimensional problems when $p$ is larger than $n$. It is known that the MLE of the precision matrix $\mathbf{\Omega}$ is $\mathbf{S}^{-1}$, where $\mathbf{S} = \mathbf{X}\mathbf{X}'/n$, with $\mathbf{X}$ being the data matrix, and $\mathbf{X}'$, the transpose of matrix $\mathbf{X}$. GLasso \cite{friedman2008sparse} uses complex optimization tools to minimize the following penalized log likelihood
\begin{equation*}
{\cal L}(\mathbf{\Omega})=\log\det(\mathbf{\Omega})-tr(\mathbf{S}\mathbf{\Omega})+\lambda_L||\mathbf{\Omega}||_{1},\quad \lambda_L >0,
\end{equation*}
where $\|\cdot\|_1$ is the absolute-value norm, $||\mathbf{A}||_1 = \sum_{ij} |a_{ij}|$, with $\mathbf{A}\equiv (a_{ij})_{i,j\in\mathcal{V}}$, and $\lambda_L$ is the regularization coefficient. The procedure results in a sparse precision matrix rather than a precise partial correlation estimation \cite{meinshausen2006high}. The GGMridge \cite{ha2014partial}, on the other hand, uses a ``ridge inverse'' $(\mathbf{S}+\lambda_R \mathbf{I}_{p})^{-1}$ in the analogy for ridge regression \cite{hoerlkennard}, which generates estimates for the partial correlation matrix 
\begin{equation*}
\mathbf{\hat{P}} = -\scale((\mathbf{S}+\lambda_R \mathbf{I}_{p})^{-1}),\quad \lambda_R>0,
\end{equation*}
where $\scale(\mathbf{A}) = \diag(\mathbf{A})^{-1/2}\mathbf{A}\diag(\mathbf{A})^{-1/2}$, and $\lambda_R$ is the restriction factor. The elements of $\mathbf{\hat{P}}$, which are in the same form as (\ref{eq1}), are tested with $\alpha_R\in (0,1)$, using empirical distributions \cite{ha2014partial}, providing significant estimates for the partial correlation, and reconstructing the underlying graph structure.

\section{Simulation Studies}\label{sim}
One of the goals of this study was to evaluate if the network structure affects the overall performance of the methods. We have chosen three graph models with different topologies: Erd\"os-R\'enyi \cite{erdos1960evolution}, Barab\'asi-Albert \cite{barabasi1999emergence}, and  Watts-Strogatz \cite{watts1998collective}. In the Erd\"os-R\'enyi network, we used a uniform distribution for the edges. The Barab\'asi-Albert graph was generated through a preferential attachment algorithm \cite{barabasi1999emergence}, and for the Watts-Strogatz network, we used the algorithm described in \cite{watts1998collective}. Parameters were chosen such that the generated graph ${\cal G} = ({\cal V}, {\cal E})$, respectively its adjacency matrix $\mathbf{A}^{*}\equiv(a_{ij}^{*})_{i,j\in\mathcal{V}}$ is sparse. 

\subsection{From adjacency to covariance matrix}

In this Section, we explain how  we construct a covariance matrix for a given adjacency matrix following
the procedure described in \cite{zhou2016}. For each generated adjacency matrix $\mathbf{A}^{*}$ we attribute random values to the non-zero elements of $\mathbf{A}^{*}$, transforming it into a positive definite covariance matrix. First, define the matrix
\begin{equation*}\label{eq:chap05_sim}
\mathbf{\Omega}_{1}=\left\{ \begin{array}{ll}
\omega_{ij} = \omega_{ji}=u_{ij}\cdot\delta_{ij}, &\text{ if } a_{ij}^{*}=1,\ i<j,a_{ij}^{*}\in \mathbf{A}^{*}\\
\omega_{ij} = \omega_{ji} = 0,& \text{ otherwise},
\end{array}\right.
\end{equation*}
where $u_{ij}$ is a uniform random variable in the interval $(0.4,0.8)$, and $\delta_{ij}$ has discrete uniform distribution with values in $\{-1,1\}$. Next, define 
$\mathbf{\Omega}_2=\mathbf{\Omega}_1+(|\nu_{min}(\mathbf{\Omega}_1)|+0.05)\mathbf{I}_p$, where $\nu_{min}(\mathbf{\Omega}_1)$ is the minimum eigenvalue of $\mathbf{\Omega}_1$, and $\mathbf{I}_p$ is a $p\times p$ identity matrix. The inverse of the precision matrix, or covariance matrix, is obtained from the transformation  $\mathbf{\Omega}^{-1} = \diag(\mathbf{u}_2)\mathbf{\Omega}_{2}^{-1}\diag(\mathbf{u}_2)$, where $\diag(\mathbf{u}_2)$ is a diagonal matrix formed by the $p$-dimensional vector $\mathbf{u}_2$, which is uniformly distributed in the interval $(1,5)$. Finally, we have a multivariate Gaussian distribution $X_{\cal V} \sim {\cal N}_{\cal V}(\mathbf{0},\mathbf{\Omega}^{-1})$ from which we obtain a sample size $n$.

\subsection{Results}

\begin{figure}
	\includegraphics[width=\textwidth]{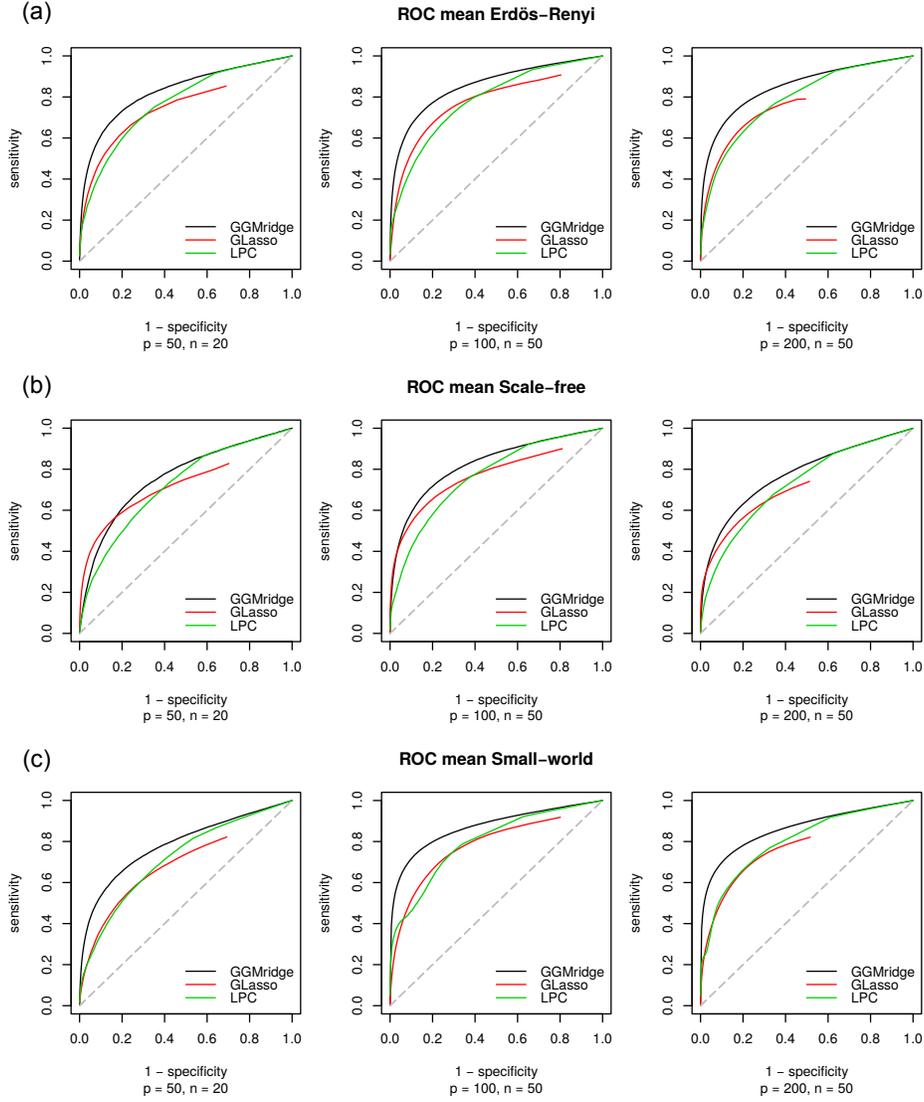}
	\caption{In the Erd\"os-R\'enyi structure (a), the GGMridge outperformed the GLasso and the LPC, with LPC having better results than the GLasso in every case, as seen in Table \ref{tab1}. In the scale-free structure (b), we can observe that the GGMridge method has a better performance than the GLasso and LPC methods. In this case, GLasso performed better, by a small margin, than LPC in the networks with $p = 50$, although the sd of LPC was smaller, and $p = 100$, with LPC obtaining better results with $p = 200$. In the small-world structure (c), LPC outperforms GLasso in every case, as confirmed in Table \ref{tab1}, although the sd is higher in the LPC. Again, the GGMridge method obtained curves with larger areas. Except for the Scale-free and small-world structure with $p = 100$, the lowest standard deviations in Table \ref{tab1} comes from the LPC.} \label{fig_ROC}
\end{figure}

The area under the ROC curve was used as a measure to compare the methods. Usually applied in binary classifiers, it describes the trade-off between the false positives fraction, which is the probability of misclassification (specificity), and the true positives, which is the probability of a correct classification (sensitivity). Therefore, the methods with larger areas under the ROC curve are considered more efficient classifiers. The values of the parameters used to plot the curves vary from the less regularized to the point where the graph is almost empty (full regularization). For the GLasso we used $\lambda_L \in \{0.001,0.006\dots,1; \text{by 0.005}\}$, in the GGMridge we used $\alpha_R \in \{0.0001,0.0011\dots,40; \text{by 0.001\}}$, and for the LPC $\alpha = \alpha_{LPC} \in \{10^{-4},\dots,0.4;\text{by 0.01} \} \cup \{0.6,0.7,0.8,0.9,1\}$ varying at the same rate. In total, we ran 300 simulations for each method with sizes: $(p=50, n=20)$, $(p=100, n=50)$, and $(p=200, n=50)$. Next, we took the average of the produced specificity and sensitivity, generating the curves in Figure \ref{fig_ROC} and the areas in Table \ref{tab1}. We can observe that, in most cases, the GGMridge  outperformed GLasso and LPC, with LPC in the scale-free, and small-world having a better performance than GLasso.

\section{Gene Expression Network}\label{gene_express}
\subsection{Motivation}\label{motv}
Gene co-expression analysis aims to identify genes whose expression differs in healthy cells in comparison with those with abnormal behavior, such as tumorous cancer cells. Since most gene co-expression data are high dimensional, with the number of genes reaching the thousands and very small sample size, performing this kind of analysis can be highly challenging. This situation inspires most of the network reconstruction methods currently being developed.

\begin{table}[t!]
	\begin{center}
		\caption{Mean of the ROC curve area by method, sample size, graph type, and standard deviation expressed in parentheses.}\label{tab1}
		\begin{tabular}{ |c|l|c|c|c| }
						\hline
			(variables, sample size) & Graph type & GGMridge & GLasso & LPC \\ 
			\hline
			& Erd\"os-R\'enyi 					& .8302(.0469)  & .7522(.0465) & .7708(.0429) \\ 
			$(p=50, n=20)$ & Watts-Strogatz 	& .7848(.0357)  & .6926(.0303) & .7171(.0301) \\ 
			& Barab\'asi-Albert 				& .7610(.0488)  & .7262(.0491) & .7206(.0446) \\
			\hline
			& Erd\"os-R\'enyi 					& .8560(.0241)  & .7794(.0243) & .7821(.0221) \\ 
			$(p=100, n=50)$& Watts-Strogatz 	& .8712(.0229)  & .7836(.0247) & .7972(.0518) \\ 
			& Barab\'asi-Albert 				& .8513(.0414)  & .7666(.0372) & .7619(.0569) \\ 
			\hline
			& Erd\"os-R\'enyi 					& .8513(.0195) 	& .7666(.0209) & .7893(.0181) \\ 
			$(p=200, n=50)$ & Watts-Strogatz 	& .8621(.0177)  & .7748(.0188) & .8012(.0166)  \\ 
			& Barab\'asi-Albert 				& .7742(.0380) 	& .7172(.0335) & .7272(.0326) \\ 
			\hline
		\end{tabular}
	\end{center}
\end{table}
Motivated by this problem, the data used in the application come from \cite{zhai2007gene}, which  contain the expression of 25,387 genes extracted from 21 tumor tissue samples from human cervical cancer \cite{ref_url1}. From these data, we selected a subset of 1,268 differentially expressed genes (DEGs), identified by \cite{mine2013gene}. Since networks cannot be compared using the same threshold, we followed heuristics provided by the biological community \cite{MYTSRD2015}, \cite{TVSYM2016}, and decided to reconstruct the networks with 3 times more edges than nodes. We used specific thresholds and regularization coefficients. In GGMridge, the values were $\lambda_R = 1$ and $\alpha_R = 0.01$, for GLasso, the value was $\lambda_L = 0.6$, and for LPC, $\alpha=0.1$ and $\alpha_{LPC}=0.02$.

\subsection{Reconstructed Networks}
\begin{figure}[t!]
	\includegraphics[width=\textwidth]{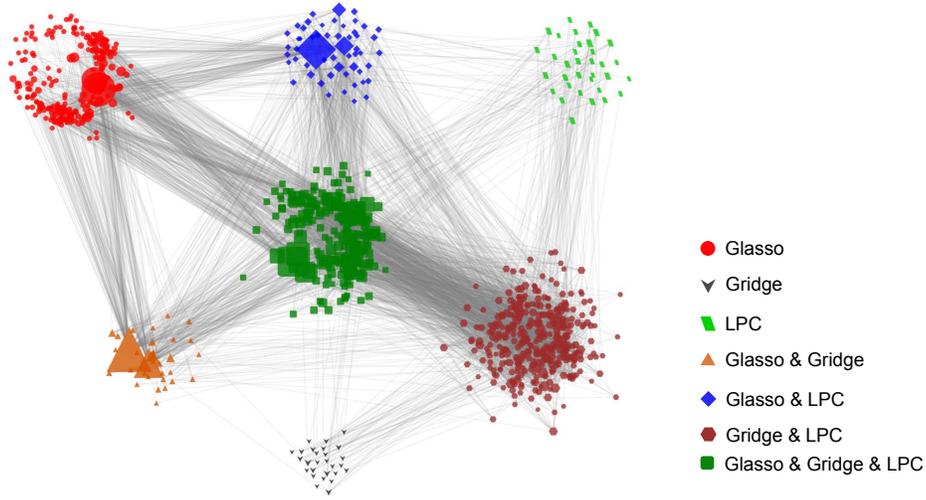}
	\caption{Genes (nodes) and expressions (edges) identified by the three methods. The 244 red circles are the nodes identified only by GLasso, the 27 gray inverted triangles are the nodes identified by GGMridge, and the 34 light green parallelograms were detected by LPC. The 33 orange triangles are the nodes identified by both GLasso and GGMridge. The 343 brown hexagons are the nodes identified by both GGMridge and LPC. The 54 blue diamonds are the nodes detected by both the GLasso and LPC, and the 227 dark green squares are the nodes identified by the three methods. The gray lines between the nodes are the edges that were identified by the methods.} 
	\label{fig1}

\end{figure}

\begin{figure}[t!]
\includegraphics[width=\textwidth]{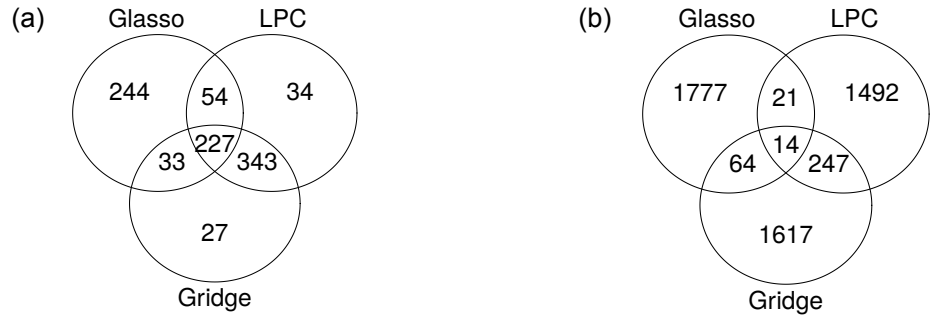}
\caption{(a) Detected nodes by method, and in (b) detected edges by method.} \label{fig2}
\end{figure}

The thresholds and regularization parameters defined in Subsection \ref{motv} results in three different networks. GGMridge generated a network with 558 nodes and 1876 edges, GLasso one with 630 nodes and 1942 edges, and LPC identified 658 nodes and 1774 edges. In Figure \ref{fig1}, generated by Cytoscape \cite{cyto}, we have the reconstructed gene co-expression network of all nodes and edges obtained in common. Figure \ref{fig2} is the Venn diagram representation of Figure \ref{fig1} and provides valuable information about the results. We can observe that LPC and GGMridge share numerous nodes, 570 in total. Including the nodes shared with GLasso, we have 570 common identifications, almost $60\%$ of the total. GLasso and LPC share 281 nodes or $27\%$, and GLasso with GGMridge share 250 nodes, $26\%$ of the total. On the other hand, the edges are not shared proportionally as the nodes, only $0.2\%$ of the edges were identified in common between all methods. While LPC and GGMridge share a considerable percentage of nodes ($4\%$), LPC and GLasso share only $0.06\%$, and GGMridge and GLasso share $1.4\%$. The degree of the nodes, which is the number of edges incident to the node, was also different. The node with a maximum degree was identified by GLasso, with size 188. In GGMridge, it was 35, and in LPC, 14. An interesting fact occurs when GGMridge and GLasso are observed together, generating a maximum degree of 188. LPC and GLasso achieve 165, and LPC and GGMridge 35. The three methods identified a maximum degree of 177, which suggests that GLasso favors nodes with higher degrees, or ``hubs''\cite{barabasi1999emergence}.


\section{Conclusion}
The simulation study provided an interesting glimpse of the three approaches mentioned above. We observed that GGMridge outperformed GLasso and LPC in all cases. Being a relatively new method, GGMridge needs more tests in order to prove its quality. LPC, although having a heuristic approach, performed better than GLasso in most cases, proving to be a viable alternative for GGM inference, without losing the partial correlation estimates.  In the application, the reconstructed networks differed from each other in significant ways. GGMridge and LPC were the methods with more common identifications regarding edges and nodes detection.
%
%
%

\end{document}